\documentclass[11pt]{article}
\usepackage[
  paper=a4paper,
  textwidth=145mm,
  textheight=232mm,
  includehead,
  includefoot,
  twoside,
  hmarginratio=1:1
]{geometry}

\usepackage{algorithm}
\usepackage{algpseudocode}
\usepackage{amsfonts}
\usepackage{amsmath}
\usepackage{amssymb}
\usepackage{amsthm}
\usepackage{anyfontsize}
\usepackage[style=ieee, backend=biber]{biblatex}
\usepackage{enumitem}
\usepackage{graphicx}
\usepackage{hyperref}
\usepackage{multirow}
\usepackage{newtxtt}
\usepackage{orcidlink}
\usepackage{subfig}

\newtheorem{theorem}{Theorem}
\newtheoremstyle{spacedef}
  {1em}   
  {1em}   
  {\itshape} 
  {}      
  {\bfseries} 
  { }     
  { }     
  {}      
\theoremstyle{spacedef}

\newcommand{\scinot}[3]{$#1 (#2\mathrm{E}{#3})$}

\providecommand{\keywords}[1]
{
  \small	
  \textbf{\textit{Keywords---}}#1
}

\usepackage{etoolbox}

\AtBeginEnvironment{algorithm}{\fontsize{10pt}{10pt}\selectfont}
\AtBeginEnvironment{algorithmic}{\fontsize{10pt}{10pt}\selectfont}
\AtBeginEnvironment{tabular}{\fontsize{10pt}{10pt}\selectfont}

\begin{document}

\title{Applicability of the Minimal Dominating Set for Influence Maximization in Multilayer Networks}

\author{
    Micha{\l} Czuba\thanks{corresponding author}\hspace{5pt}\thanks{are with Department of Artificial Intelligence, Wroc{\l}aw University of Science and Technology, 27 wybrze{\.z}e Wyspia{\'n}skiego st, 50-370 Wroc{\l}aw, Poland, e-mails: {\tt \{michal.czuba, piotr.brodka\}@pwr.edu.pl}}\hspace{5pt}\thanks{are with Complex Adaptive Systems Lab, Data Science Institute, School of Computer Science, University of Technology Sydney, Ultimo NSW 2007, Australia, e-mails: {\tt \{mingshan.jia, katarzyna.musial-gabrys\}@uts.edu.au}}\hspace{5pt}\orcidlink{0000-0001-8652-3678}
    Mingshan Jia\footnotemark[3]\hspace{5pt}\orcidlink{0000-0001-8378-3407}
    Piotr Br{\'o}dka\footnotemark[2]\hspace{5pt}\footnotemark[3]\hspace{5pt}\orcidlink{0000-0002-6474-0089}
    Katarzyna Musial\footnotemark[3]\hspace{5pt}\orcidlink{0000-0001-6038-7647}
}

\maketitle

\begin{abstract}
    The minimal dominating set (MDS) is a well-established concept in network controllability and has been successfully applied in various domains, including sensor placement, network resilience, and epidemic containment. In this study, we adapt the local-improvement MDS routine and explore its potential for enhancing seed selection for influence maximization in multilayer networks. We employ the Linear Threshold Model (LTM), which provides an intuitive representation of influence spread or opinion dynamics by accounting for the accumulation of peer influence. To ensure interpretability, we utilize rank-refining seed selection methods, with the results further filtered with MDS. Our findings reveal that incorporating MDS into the seed selection process improves spread only within a specific range of situations. Notably, the improvement is observed for larger seed set budgets, lower activation thresholds, and when an $AND$ strategy is used to aggregate influence across network layers. This scenario reflects situations where an individual does not require the majority of their acquaintances to hold a target opinion, but must be influenced across all social circles.
\end{abstract}


\section{Introduction}
\label{sec:intro}

Traditionally, spreading processes, including influence diffusion, were studied on simple, undirected, and unlabelled networks. However, these models fail to represent real-world relationships realistically. With time, more and more complex graph representations were developed, and eventually, a multilayer network concept was proposed. It offers a more nuanced depiction of social interactions than classical graphs. This model captures diverse types of connections --- such as communication, professional, family, or social ties, by distinguishing between different layers (i.e., types of relationships)~\cite{dickison2016multilayer, kivela2014multilayer}. Hence, incorporating such networks into research on diffusion increases the accuracy of modelling.

One example of spreading processes is influence spread, a key challenge in social network analysis with applications in areas such as viral marketing, information diffusion, and public health interventions~\cite{salehi2015spreading}. The process is often analyzed through cascade dynamics using mechanisms such as the Linear Threshold Model (LTM) and the Independent Cascade Model (ICM). LTM employs a threshold-based activation rule, where a node becomes active only if the total weight of its active neighbours exceeds a predefined threshold~\cite{granovetter1978threshold, watts2002simple}. In contrast, ICM activates nodes probabilistically through independent trials along each edge~\cite{goldenberg2001talk}. While ICM has been extended to multilayer settings~\cite{erlandsson2018seed, brodka2021sequential, lin2023tensor, achour2024theoretical}, its limited flexibility in capturing heterogeneous link influences makes LTM a more suitable choice for modelling opinion dynamics in a multilayer environment~\cite{yaugan2012analysis, zhong2022mltm}. Researchers developed various LTM extensions for multilayer networks, e.g., the approach presented in~\cite{yaugan2012analysis} incorporates content-dependent parameters to differentiate link types, while another one~\cite{zhong2022mltm} introduces layer-specific activation thresholds and heterogeneous activation rules. Given a cascade model, the next key step of influence maximization is to identify seed agents that achieve the widest spread with minimal resources~\cite{kempe2003maximizing}. Numerous methods have been proposed for seed selection in multilayer networks and, as categorized in~\cite{singh2022influence}, can be grouped into heuristic techniques~\cite{venkatakrishna2023kpp_shell}, simulation-based approaches~\cite{chen2020maximizing}, and hybrid methods incorporating machine learning~\cite{yuan2024gbim, shu2025node}.

Influence maximization can be viewed as a softer version of control, which motivates this study's exploration of incorporating concepts from control theory to assess their applicability in the realm of influence spread. Control theory in network science focuses on strategies to transition a system from one state to another within a defined number of steps~\cite{liu2011controllability}. To achieve this, control methods have been developed that identify a minimal set of driver nodes capable of steering the network toward the desired state~\cite{liu2016control}. For example, a study by~\cite{doostmohammadian2019minimal} discusses the identification of a minimal set of driver nodes for structural controllability in large-scale dynamical systems. Some researchers have examined the impact of multilayer networks on the controllability of these structures~\cite{jiang2022controllability}. However, real social networks often do not exhibit deterministic behaviours, making strict network control difficult.

In this paper, we aim to bridge concepts from control theory with influence maximization techniques, which have been largely independent of each other. We address this through an exploratory study that employs the Minimal Dominating Set (MDS) in the seed selection process for multilayer networks. We explore the viability and potential advantages of using MDS to maximize influence. Previous research has shown that MDS is an effective seed selection method in single-layer networks~\cite{sadaf2022maximising, sadaf2024bridge}, but its performance in multilayer networks remains largely unexplored. Note that the primary goal of this study is not to propose a new seed selection method but to assess the effectiveness and usefulness of MDS in improving existing methods.

Our study demonstrates that incorporating MDS into the seed selection process enhances diffusion, but this improvement occurs only within a specific range of spreading parameters. The benefit is most noticeable when the activation thresholds are lower, the $AND$ strategy is used to aggregate influence across network layers, and the seed set budget is larger. Additionally, the enhancement is particularly pronounced in networks resembling scale-free graphs. This scenario is especially relevant when an individual needs to be influenced across all social circles without requiring the majority of their acquaintances to share the target opinion. An example might be a presidential election with highly polarized candidates; we may not know the opinions of all our friends and family because not everyone wants to share them (for instance, they are afraid of what others would think), or we prefer not to discuss politics. However, we usually know the opinions of a few people in each of our social circles, and if the opinions of all (or most) of them are the same, we are more willing to adopt it as our view.

The article is organized as follows. In Sec.~\ref{sec:mds_in_mln}, we present a formal definition of MDS, review state-of-the-art methods for obtaining it, and introduce our adaptation of the local-improvement algorithm to the context of multilayer networks. Sec.~\ref{sec:setup} outlines the experimental setup used in the study, including metrics and the evaluated parameter space. The results of the analysis are discussed in Sec.~\ref{sec:results}, where we examine the properties of the MDSs obtained for the examined graphs, the similarity of seed sets drawn with and without incorporating MDS in the seed selection process, and the effectiveness of that in maximizing influence with a particular attention to scale-free networks. Sec.~\ref{sec:conclusions} concludes the work.

\section{MDS in Multilayer Networks}
\label{sec:mds_in_mln}

MDS is the minimal set of ``driver agents'' that are capable of controlling the entire network. The problem of identifying it is well known to be NP-hard~\cite{karp2010reducibility}, and numerous approaches have been proposed to solve it efficiently. In this section, we introduce this problem, provide a review of related works, and introduce an extension of the local improvement algorithm to the multilayer setting.

\subsection{Multilayer networks}

First, we clarify how we conceptualize multilayer networks. Specifically, we adopt the framework introduced by~\cite{dickison2016multilayer, kivela2014multilayer}. To ensure consistency in notation, we restate it as follows:

\begin{theorem}[Multilayer Network]
    \label{def:multilayer_net}
    A multilayer network can be described as a tuple $M = (A,L,V,E)$ consisting of the following sets:
    \begin{itemize}[noitemsep, topsep=0pt]
        \item actors $A=\{a_1, a_2, \dots\}$,
        \item layers $L=\{l_1, l_2, \dots\}$, 
        \item nodes $V=\{v_1^{1}, v_1^{2}, \dots, v_2^{1}, v_2^{2}, \dots\}: V \subseteq A \times L$,
        \item edges $E=\{(v_1^{1}, v_2^{1}), \dots, (v_1^{2}, v_2^{2}), \dots\}: (v_1^{1}, v_2^{2}) \notin E \land (v_1^{1}, v_2^{1}) \equiv (v_2^{1}, v_1^{1})$.
    \end{itemize}
\end{theorem}

In other words, a multilayer network is a coupled set of single-layer networks, where each actor is represented in at most every layer by a corresponding node ($V \subseteq A \times L$), e.g., the representation of actor $a_1$ in layer $l_2$ is denoted as $v_1^{2}$. Another property of this model is that interlayer edges are not allowed, meaning links can only exist within the same layer: $(v_1^{1}, v_2^{2}) \notin E$. Additionally, to simplify the analysis, edges are considered undirected: $(v_1^{1}, v_2^{1}) \equiv (v_2^{1}, v_1^{1})$ and unweighted.

\subsection{Dominating sets}

Having introduced the network model utilized in this work, we are able to provide the necessary definitions of dominating sets, extending the foundational concepts introduced in~\cite{nacher2012dominating} to the multilayer scenario.

\begin{theorem}[Dominating Set]\label{def:ds}
    A set of actors $D'$ dominates the multilayer network $M$ if for every actor $a \in A$, either $a$ belongs to $D'$, or for each layer where $a$ is represented, the corresponding node of $a$ is adjacent to a node that represents an actor in $D'$.
\end{theorem}

\begin{theorem}[Minimum Dominating Set]\label{def:m-umds}
    Let $\mathcal{D}$ be a family of dominating sets of the multilayer network $M$: $\mathcal{D} \subseteq powerset(A)$. A Minimum Dominating Set is a set $D: D \in \mathcal{D}$, and $|D| = \arg\min_{D' \in \mathcal{D}} |D'|$, i.e., $D$ is the smallest dominating set among all dominating sets in $\mathcal{D}$.
\end{theorem}

\begin{theorem}[Minimal Dominating Set]\label{def:m-alds}
    A set of actors $\hat{D}$ is called a minimal dominating set if it satisfies Def.~\ref{def:ds} and none of its proper subsets is a dominating set.
\end{theorem}

As stated above, many dominating sets can be obtained for a particular network, and in the edge case, the set of all actors also constitutes a dominating set. Another implication of the aforementioned definitions is that multiple minimum and minimal dominating sets may exist for a given multilayer network. Finally, both Def.\ref{def:ds} and Def.\ref{def:m-umds} are consistent with the original concept of the minimum dominating set introduced for classic (i.e., static, undirected, and single-layer) graphs~\cite{nacher2012dominating}, when they are applied to a single-layer network (a special case of a multilayer network with $|L|=1$). An additional important remark is that following~\cite{casado2023iterated}, we distinguish the minim\textbf{um} dominating set (Def.\ref{def:m-umds}) from the minim\textbf{al} dominating set (Def.\ref{def:m-alds}), as the latter can be larger than the former. This distinction is particularly important in networks for which finding $D$ within a reasonable amount of time is infeasible.

\subsection{Example}

\begin{figure}[ht]
	\centering
	\includegraphics[width=.35\linewidth]{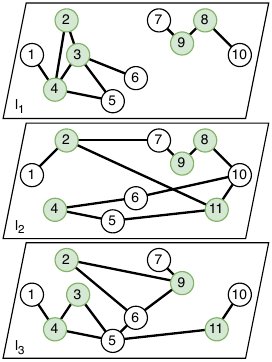}
	\caption{An example of a multilayer network with three types of relations ($l_1$, $l_2$, $l_3$). Nodes which represent actors that belong to the MDS are highlighted in green colour ($\hat{D}={a_2, a_3, a_4, a_8, a_9, a_{11}}$).
    }
    \label{fig:toy_mds}
\end{figure}

To illustrate both the MDS and an example of a multilayer network, we present Fig.~\ref{fig:toy_mds}. Note that even a small structural change can affect the MDS. For instance, adding an edge between nodes $n_9$ and $n_6$ in layer $l_1$ would allow actor $a_3$ to be removed from the MDS. The set shown in Fig.~\ref{fig:toy_mds} satisfies both Def.~\ref{def:ds} and Def.~\ref{def:m-alds}, but not Def.~\ref{def:m-umds}, as there exist dominating sets of smaller size dominating the network. Specifically, the minimum dominating sets for the given network have size~$5$, and eight such sets exist, e.g., $D = \{a_2, a_3, a_4, a_9, a_{10}\}$. Although the example is a simple toy case, it has been deliberately chosen to demonstrate that computing $D$ is non-trivial. The algorithm employed in this study yields $\hat{D}$, which serves as an approximation of $D$ in terms of size.

\subsection{Related work}

The problem of finding MDS was originally formulated for single-layer graphs and has since been extensively studied in the literature. The classic greedy heuristic~\cite{chvatal1979} serves as a foundational method and, in addition, offers a theoretical approximation guarantee. However, in practical scenarios, it often overestimates the true MDS size due to its myopic nature, leading to suboptimal solutions. Metaheuristic approaches have been widely explored to improve upon traditional greedy approaches. The Ant Colony Optimization Algorithm hybridized with a Local Search (ACO-LS) leverages pheromone trails for guiding the search process while refining solutions through local search~\cite{potluri2011two}. Building on this, an extended version, ACO-PP-LS, incorporates a pre-processing phase to generate initial solutions using a greedy heuristic, thereby accelerating convergence~\cite{potluri2013hybrid}. Beyond ACO-based methods, order-based Randomized Local Search has been proposed, representing solutions as permutations of vertices and applying a perturbation mechanism to refine dominating sets iteratively~\cite{chalupa2018order}. While RLS improves upon many existing methods, it remains computationally demanding and is sensitive to the choice of its perturbation operator. The Iterated Greedy (IG) algorithm has been introduced as an efficient and effective approach for finding MDS~\cite{casado2023iterated} to address these limitations. By iteratively destructing and reconstructing solutions while incorporating a specialized local search, IG achieves a superior balance between diversification and intensification. Empirical evaluations demonstrate that IG not only outperforms ACO-based methods and RLS in solution quality but also significantly reduces computational effort, making it a state-of-the-art approach.

Since most existing methods focus on single-layer networks, the problem of finding the MDS in multilayer networks remains largely unexplored. One of the most recent approaches, FAST-MDSM~\cite{nacher2019finding}, employs integer linear programming (ILP) combined with graph reduction heuristics to identify a minimal set of driver nodes across multiple interconnected layers. However, its reliance on ILP renders it computationally expensive and less adaptable to dense graphs. Notably, FAST-MDSM was designed for scale-free networks and has been primarily tested on sparse biological networks, limiting its applicability to more general settings. In this work, inspired by the IG algorithm~\cite{casado2023iterated}, originally developed for single-layer networks, we propose a new approach for efficiently solving the MDS problem in multilayer networks.

\subsection{Proposed algorithm}

An extension of the local improvement routine from~\cite{casado2023iterated} to multilayer networks is presented as Alg.~\ref{alg:driver_actor_selection}. As an input, it accepts a multilayer network $M$ and an initial dominating set $D'$ obtained using a greedy routine, as described in~\cite{czuba2024networkdiffusion}. Note that $D'$ is likely to be only subminimal, as it is obtained using a relatively simple heuristic; thus, the subsequent operations aim to reduce its size while ensuring that it continues to satisfy Def.~\ref{def:m-alds}.

It is important to understand the quality of the initial solution before applying the local improvement algorithm. To assess this, we refer to~\cite{sanchis2002experimentalds}, where the authors evaluated a degree-based greedy heuristic for finding MDS in single-layer networks with $400 \leq |V| \leq 800$, for which the sizes of the minimum dominating sets were known. Experiments conducted on 60 networks revealed that the median gap between a greedily obtained MDS and the optimum is around $35\%$, decreasing to approximately $10\%$ for graphs with a density of $0.5$. Referring these results to the multilayer setting, which is more complex, we estimate that the initial solution provided to the local improvement routine is no closer than $35\%$ to the optimal value.

First (l. 1--3), we compute a domination map (\textit{ComputeDomination}) that, for each layer of $M$, associates nodes representing actors from $D'$ with sets of nodes they dominate. We then initialize a flag to control the termination of the routine and assign $D'$ as the current best solution. The next step (l. 4--27) consists of a loop that explores alternative configurations of dominating sets derived from $D$ in an attempt to identify the smallest one. The evaluation of actors from the initial solution, as well as the exploration of potential replacements for them, is carried out in a random order with the \textit{Shuffle} function. That is aimed to ensure that the algorithm produces different MDSs for the same graph across subsequent runs. This approach helps mitigate the risk that the results depend on a single instance of the MDS drawn for a given network.

Specifically, the \textit{for} loop (l. 6) iterates over the actors in $\hat{D}$ to assess whether replacing a particular actor $a_{\hat{D}}$ leads to a reduction in its size. To achieve this, we identify a set of candidate replacements $C$ for each $a_{\hat{D}}$ using the auxiliary function \textit{FindReplacementCandidates}. For each actor $a_C \in C$, we construct a new set $\hat{D}_{\text{new}}$ by substituting $a_{\hat{D}}$ with $a_C$ in $\hat{D}$. Then, using \textit{IsFeasible}, $\hat{D}_{\text{new}}$ is examined to determine if it meets the criteria to dominate $M$. If the set remains feasible, we attempt to greedily minimize its size by iteratively removing redundant actors. If the pruned set $\hat{D}_{\text{reduced}}$ is not smaller than the current best solution, we continue iterating through $C$. Otherwise, we update $\hat{D}$ with $\hat{D}_{\text{reduced}}$, recompute the domination map for the updated $\hat{D}$, and reset the $improved$ flag. At this point, we restart the outer loop, attempting further refinement of the dominating set. The iteration continues until no further refinements can be made. At this stage, the algorithm terminates and returns $\hat{D}$, which is locally minimal in size.

\begin{algorithm}[ht]
    \caption{MDS with Local Improvement for Multilayer Networks}
    \label{alg:driver_actor_selection}
    \begin{algorithmic}[1]
        \Require $M = (A, L, V, E)$ \Comment{multilayer network}
        \Require $D'$ \Comment{initial dominating set}
        \State $domination \gets \text{ComputeDomination}(M, D')$
        \State $improved \gets \text{True}$
        \State $\hat{D} \gets D'$
        \While{$improved$}
            \State $improved \gets \text{False}$
            \For{$a_{\hat{D}} \in \text{Shuffle}(\hat{D})$}
                \State $C \gets \text{FindReplacementCandidates}(M, a_{\hat{D}}, \hat{D}, domin)$
                \For{$a_C \in \text{Shuffle}(C)$}
                    \State $\hat{D}_{\text{old}} \gets \hat{D}$
                    \State $\hat{D}_{\text{new}} \gets (\hat{D} \setminus \{a_{\hat{D}}\}) \cup \{a_C\}$
                    \If{$\text{IsFeasible}(M, \hat{D}_{\text{new}})$}
                        \State $\hat{D}_{\text{reduced}} \gets \text{RemoveRedundantVertices}(M, \hat{D}_{\text{new}})$
                        \If{$|\hat{D}_{\text{reduced}}| < |\hat{D}_{\text{old}}|$}
                            \State $\hat{D} \gets \hat{D}_{\text{reduced}}$
                            \State $domination \gets \text{ComputeDomination}(M, \hat{D})$
                            \State $improved \gets \text{True}$
                            \State \textbf{break}
                        \Else
                            \State $\hat{D} \gets \hat{D}_{\text{old}}$
                        \EndIf
                    \EndIf
                \EndFor
                \If{$improved$}
                    \State \textbf{break}
                \EndIf
            \EndFor
        \EndWhile
    \State \Return $\hat{D}$ \Comment{improved dominating set}
    \end{algorithmic}
\end{algorithm}

The efficiency of the proposed algorithm also plays an important role and can be estimated by assessing its complexity. First, since Alg.~\ref{alg:driver_actor_selection} builds on the initial solution, the time complexity of the greedy approach, $T_{\text{greedy}} = O(L\,|A|^{2}\,|D'|)$, is taken as a reference. Once the local improvement phase is added, the algorithm's complexity is dictated by the search for replacements and the subsequent redundancy removal. This yields a total time complexity of $T_{\text{greedy+LI}} = O(L\,|A|^{2}\,|D'|^{5})$ for dense graphs. Therefore, the greedy algorithm with local improvement is costlier than the naive greedy approach by a factor of $|D'|^{4}$. While this worst-case complexity appears computationally expensive, the algorithm's runtime in practice is often significantly lower. This practical efficiency stems from two key factors. First, the number of successful improvements found by the outer \textit{while} loop is typically a manageable percentage of $|D'|$. Second, and more importantly, the number of candidate nodes $C$ that can replace a node $a_{\hat{D}}$ in the dominating set is also moderate. The constraint that a candidate must dominate all nodes exclusively covered by $a_{\hat{D}}$ is highly restrictive. As a result, the inner loop rarely iterates over all $|A|$ vertices; it usually inspects just a handful of viable replacements, making the local improvement phase tractable for small- to medium-sized graphs despite the pessimistic worst-case bound.

\section{Experimental Setup}\label{sec:setup}

In this section, we describe the simulation setup, including the dataset and the diffusion model applied. We then present the baseline seed selection methods and explain how MDS was incorporated into them. We also outline metrics used to assess diffusion effectiveness, an evaluated parameter space, and the computational resources employed.

\subsection{Dataset}\label{subsec:dataset}

The efficiency of MDS depends on the structure of the underlying network. Furthermore, as mentioned in Sec.~\ref{sec:intro}, dynamics in multilayer networks often differ significantly from those observed in single-layer networks. Consequently, selecting appropriate data for experiments is crucial. To ensure a comprehensive evaluation, we utilize both real-world and artificially generated graphs. A summary of the networks used in this study, with a brief description of the domain context in which each was constructed, is provided in Tab.~\ref{tab:networks_eda}. All artificial networks were generated using the \texttt{multinet} library~\cite{magnani2021analysis}. The real-world networks come from different domains, but all represent human-based relationships.

\begin{table}[ht]
    \caption{Networks used in the experiments, with their parameters and corresponding domain context.
    }
    \addtolength{\tabcolsep}{-.38em}
    \begin{tabular}{llrrrrp{7.2cm}}
    Type & Name & Layers & Actors & Nodes & Edges & Note \\ \hline \hline
    \multirow{3}*{\rotatebox{90}{E-R}} & er-2 & 2 & 1,000 & 2,000 & 5,459 & Erd\H{o}s-R\'{e}nyi network~\cite{er_model} generated with~\cite{magnani2021analysis}. \\
    & er-3 & 3 & 1,000 & 3,000 & 7,136 & Erd\H{o}s-R\'{e}nyi network~\cite{er_model} generated with~\cite{magnani2021analysis}. \\
    & er-5 & 5 & 1,000 & 5,000 & 15,109 & Erd\H{o}s-R\'{e}nyi network~\cite{er_model} generated with~\cite{magnani2021analysis}. \\ \hline
    \multirow{3}*{\rotatebox{90}{S-F}} & sf-2 & 2 & 1,000 & 2,000 & 4,223 & scale-free network~\cite{sf_model} generated with~\cite{magnani2021analysis}. \\
    & sf-3 & 3 & 1,000 & 3,000 & 5,010 & scale-free network~\cite{sf_model} generated with~\cite{magnani2021analysis}. \\
    & sf-5 & 5 & 1,000 & 5,000 & 10,181 & scale-free network~\cite{sf_model} generated with~\cite{magnani2021analysis}. \\
    \multirow{12}*{\rotatebox{90}{REAL}} & arxiv & 13 & 14,065 & 26,796 & 59,026 & Coauthorship network obtained from articles published on the "arXiv" repository~\cite{dedomenico2015arxiv}. \\
    & aucs & 5 & 61 & 224 & 620 & Interactions between employees of \textbf{A}arhus \textbf{U}nversity, Dep. of \textbf{C}omputer \textbf{S}cience~\cite{rossi2015towards}. \\
    & ckmp & 3 & 241 & 674 & 1,370 & A network depicting diffusion of innovations among physicians~\cite{coleman1957ckmp}. \\
    & lazega & 3 & 71 & 212 & 1,659 & A network of various types of interactions between staff of a law corporation~\cite{snijders2006lazega}. \\
    & l2-course & 2 & 41 & 82 & 297 & A first snapshot of the network built from interactions between students during a three-month-long abroad language course in Arabic~\cite{paradowski2024l2_course}. \\
    & timik & 3 & 61,702 & 102,247 & 881,676 & A graph of interactions between users of the virtual world platform for teenagers~\cite{jankowski2017timik}. \\ \hline
    \end{tabular}
    \label{tab:networks_eda}
\end{table}

\vspace{-1em}

\subsection{Spreading model}\label{subsec:ltm}

We use the Multilayer Linear Threshold Model (MLTM) introduced in~\cite{czuba2024rankrefininginfmaxmln}. It simulates diffusion in a multilayer network where actors are the subject of spreading. The model uses two thresholds: $\mu$, the traditional threshold from the classic LTM, which dictates that a node receives positive input when the sum of weighted influences from its neighbours exceeds this value, and $\delta$ (referred to as the protocol), which governs how inputs from nodes representing an actor across layers are aggregated. When the protocol threshold is exceeded, the actor (and all nodes representing it) becomes active; otherwise, the actor remains inactive, even if some of the corresponding nodes in the layers receive positive input.

At this point, it is worth clarifying the rationale behind introducing the double-threshold mechanism. Unlike studies such as~\cite{ke2024opinion, nowak2020symmetricaltm, abella2024orderingdynamics}, which modify baseline spreading models to more accurately simulate specific phenomena (e.g., to allow for bidirectional binary opinion changes~\cite{ke2024opinion}), the protocol function here serves solely as a fusion mechanism to determine the state of an actor given the states of nodes representing it. Thus, this modification should be regarded as an extension of the classic LTM to a different network model --- multilayer networks (in a manner analogous to, e.g., hypergraphs~\cite{chen2025globalcascades}), rather than as a change aimed at incorporating additional social mechanisms.

In this work, we specifically use two edge cases for the protocol function of the MLTM: $AND$, where positive input at all layers is required to activate an actor, and $OR$, where positive input at just one layer is sufficient for activation. As in~\cite{czuba2024rankrefininginfmaxmln}, we will denote the set of active actors at time step $t$ as $S_t$ (with the seed set denoted as $S_0$). Moreover, to clarify the definitions of the performance metrics (see Sec.~\ref{subsec:metrics}) employed in this study, we introduce an additional function as in Def.~\ref{def:spreading_mltm}.

\begin{theorem}[Spreading Dynamics under MLTM]
    Let $Y$ be a function to measure the spreading dynamics in a network $M$ under the Multilayer Linear Threshold Model:
    \[
    Y(M, t) = \sum_{a \in A} y_a(t),
    \]
    where $y_a(t) \in \{0, 1\}$ denotes the state of actor $a$ at time step $t$.
    \label{def:spreading_mltm}
\end{theorem}

\subsection{Seed selection methods}

\subsubsection{Basic heuristics}\label{subsubsec:basic_ssm}

We employed five seed selection methods acting as a baseline for the further comparison with MDS: Degree Centrality (denoted as \textit{deg-c}), Degree Discount~\cite{chen2009degree_discount} (\textit{deg-cd}), One-Hop Neighbourhood Size~\cite{magnani2011ml} (\textit{nghb-1s}), Neighbourhood Size Discount~\cite{czuba2024rankrefininginfmaxmln} (\textit{nghb-sd}), and Random Choice (\textit{random}). These methods belong to the class of rank-refining techniques, and their selection was motivated by their ability to generate a complete ranking of all actors within the network, ensuring compatibility with the application of MDS. We use each of these methods as described in~\cite{czuba2024rankrefininginfmaxmln}.

\subsubsection{Sorting with MDS}\label{subsubsec:mds_ssm}

To evaluate the effectiveness of MDS, all of the methods above were modified to include additional steps for filtering out actors that do not belong to $\hat{D}$ (see Alg.~\ref{alg:seed_selection_mds}). With such a setting, we were able to compare spreading efficiency on two versions of a particular seed selection method and assess the usefulness of MDS.

\begin{algorithm}[ht]
    \caption{Seed Selection with MDS}
    \label{alg:seed_selection_mds}
    \begin{algorithmic}[1]
        \Require $M = (A, L, V, E)$ \Comment{multilayer network}
        \Require $s$ \Comment{seed set budget}
        \Require $\phi(M) \to A_{\phi}$ \Comment{rank-refining seed selection heuristic}
        \Require $\kappa(M) \to \hat{D}: \hat{D} \subseteq A$ \Comment{function returning MDS of $M$}
        \State $\hat{D} \gets \kappa(M)$
        \If{$|\hat{D}| < s$}
            \State \Return \texttt{NaN}
        \EndIf
        \State $A_{\phi} \gets \phi(M)$
        \State $S \gets \emptyset$ \Comment{seed set}
        \For{$a \in A_{\phi}$}
            \If{$a \in \hat{D}$}
                \State $S \gets S \cup \{a\}$
            \EndIf
            \If{$|S| = s$}
                \State \textbf{break}
            \EndIf
        \EndFor
        \State \Return $S$
    \end{algorithmic}
\end{algorithm}

Selecting a seed set with MDS consists of several steps. First, the minimal dominating set $\hat{D}$ is obtained using Alg.~\ref{alg:driver_actor_selection}. Simultaneously, a sorted list of all actors, $A_{\phi}$, is created based on a particular heuristic $\phi$ listed in Sec.~\ref{subsubsec:basic_ssm}. The final seed set consists of the first $\lfloor s \cdot |A| \rfloor$ actors from $A_{\phi}$ that also belong to $\hat{D}$. However, particularly for large budgets, $|\hat{D}|$ can be smaller than the maximal number of actors that can be used as a seed set. In such cases, we exclude the simulation from the evaluation process since assessing the effectiveness of MDS is impossible.

It is good to note that we slightly modified Alg.~\ref{alg:driver_actor_selection} to address scalability issues. Specifically, for the largest evaluated network, we observed that the algorithm could not be completed within a reasonable time. To mitigate this, we introduced a timeout threshold for refining the initial solution. It was set to 5 minutes per 1000 actors, leading to a maximum runtime of over 5 hours per MDS search in \textit{timik}.

\subsection{Metrics}\label{subsec:metrics}

To assess the seed set quality, two metrics were used: \textit{Gain} ($\Gamma$) for the overall effectiveness and the \textit{Area under Curve} ($\Lambda$) for the dynamics of the diffusion. Their formulas are presented in Tab.~\ref{tab:metrics}. This study evaluates whether MDS improves seed sets; thus, higher $\Gamma$ and $ \Lambda $ indicate a more effective seed set.

\begin{table}[ht!]
    \caption{Spreading metrics employed in the study.
    }
    \addtolength{\tabcolsep}{-.20em}
    \begin{tabular}{llp{9.8cm}}
    Symbol & Name  & Description \\ \hline \hline
    $\Lambda$ & Area under Curve & Area under a normalized curve of spreading dynamics: \hspace{10em} $\Lambda = \int_{0}^{1}Y_{norm}(M, t) dt$, $\Lambda \in [0, 1]$ \\ \hline
    $\Gamma$ & Gain & Total network coverage achieved by a seed set $S_0$ as a proportion of the actors activated during the diffusion and the total number of activatable actors: $\Gamma = \frac{|S_{\infty} - S_{0}|}{|A - S_{0}|}, \Gamma \in [0, 1]$. \\
    \end{tabular}
    \label{tab:metrics}
\end{table}

\begin{figure}[ht]
	\centering
	\includegraphics[width=.48\linewidth]{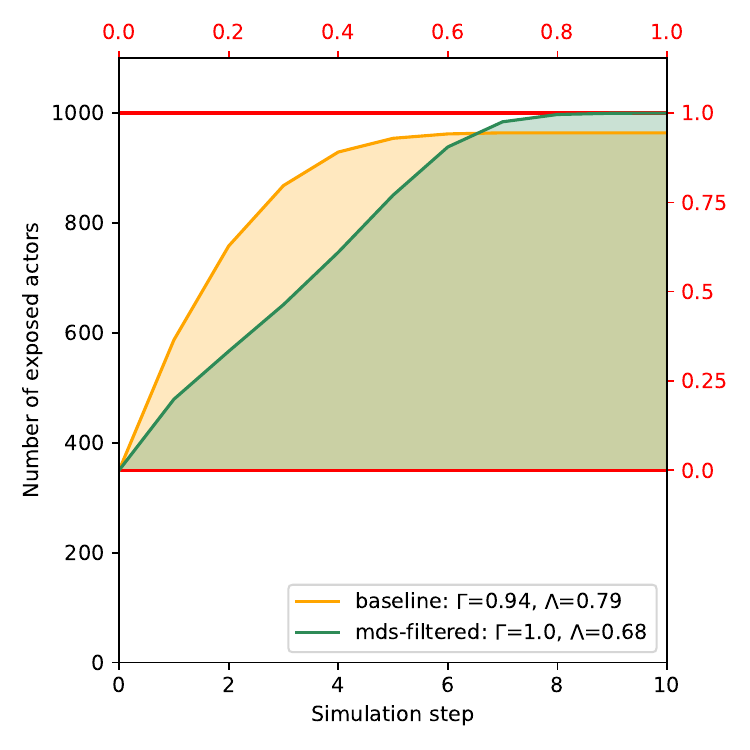}
	\caption{An example of $\Gamma$ and $\Lambda$ attained by the diffusion process within the \textit{er-3} network under MLTM with $\mu = 0.4$, $\delta=AND$, initiated from a seed set selected using \textit{nghb-sd} with $s=35$. The black scale represents real values, whereas the red scale is used for computing $\Lambda$. Red lines indicate the number of pre-activated actors (i.e., those belonging to $S_0$) and the total number of actors in the network.
    }
    \label{fig:metrics}
\end{figure}

A way how the metrics are calculated is presented in Fig.~\ref{fig:metrics}. In that example, we observe that the seed set obtained with an additional MDS filtering achieves better diffusion performance in terms of the final number of activations ($\Gamma$), yet it impairs the diffusion dynamics ($\Lambda$).

\subsection{Parameter space}

To ensure the feasibility of the experiments, we selected the parameter space based on initial trials. That allowed us to exclude cases where diffusion always failed to initiate or saturated immediately. Notably, we assigned different budget sizes for the two protocols and considered node activation thresholds ranging from $0.1$ to $0.9$ in increments of $0.1$. Note that we did not introduce heterogeneity into the model --- all nodes were assigned the same threshold value. Although spreading under the MLTM is deterministic, there are many possible combinations of the MDS for each graph. To account for this, we conducted 30 independent runs for each set of evaluated parameters, drawing a new MDS in each case.

\begin{table}[ht]
    \centering
    \caption{Diffusion parameters evaluated in the study.
    }
    \addtolength{\tabcolsep}{1.7em}
    \begin{tabular}{lcc}
    Parameter & \multicolumn{2}{c}{Values range} \\ \hline \hline
    Protocol ($\delta$) & $OR$ & $AND$ \\
    Budget ($s$) [\% of $|A|$] & $\{5, 10, 15, 20, 25\}$ & $\{15, 20, 25, 30, 35\}$ \\
    Treshold ($\mu$) & \multicolumn{2}{c}{$\{0.1, 0.2, 0.3, 0.4, 0.5, 0.6, 0.7, 0.8, 0.9\}$} \\
    Seed selection method ($\phi$) & \multicolumn{2}{l}{\{deg-c, deg-c-d, nghb-1s, nghb-sd, random\}} \\
    MDS repetitions & \multicolumn{2}{c}{$30$} \\ 
    \end{tabular}
    \label{tab:parameters}
\end{table}

In total, each seed selection method was evaluated across $90$ different MLTM parameter combinations, resulting in $10,800$ scheduled experiments. Tab.~\ref{tab:parameters} outlines the evaluated parameter space. It is worth noting that, as a follow-up to this main study, a detailed evaluation on an additional scale-free network cohort was conducted (see Sec.~\ref{subsec:sf_followup}). However, to maintain consistency, we do not include the parameters of those experiments here, as they constitute a subset of those listed in Tab.~\ref{tab:parameters}.

\subsection{Implementation}

The experiments were implemented in Python 3.12, utilizing spreading models and seed selection methods from the \texttt{network-diffusion} library~\cite{czuba2024networkdiffusion}, version 0.17. The code was designed to ensure full reproducibility, even for stochastic simulations. All experiments were conducted on a workstation running Ubuntu 20.04.4 LTS with kernel 6.5.3-arch1-1, equipped with 376 GB of memory and an Intel(R) Xeon(R) Gold 6238 CPU @ 2.10 GHz with an x86\_64 architecture. The source code,  data, and results are available at: \url{https://github.com/network-science-lab/infmax-mds-ltm-mln}.

\section{Results Analysis}
\label{sec:results}

We conduct the results analysis in four steps. First, we examine the MDSs obtained for each network to assess their variability with respect to the network structural properties. Next, we compare the seed sets used to initiate diffusion, focusing on the differences between those selected by the baseline heuristics and their MDS-filtered versions. Finally, based on the spreading metrics (Sec.~\ref{subsec:metrics}), we evaluate the impact of incorporating MDS into the seed selection process. As we pay particular attention to the scale-free model, we also conduct additional experiments to ensure that the results regarding the applicability of MDS in these networks remain consistent despite variations in their configurational parameters.

\subsection{Characteristics of MDSs}

We begin the analysis by examining the MDSs obtained for each network during the experiments to assess their variability. The results are presented in Tab.~\ref{tab:mds_networks}. The first column contains the size range of MDSs obtained during the experiments, followed by the average MDS size with standard deviation, both normalized by network size. Next, we report the number of unique MDSs out of the total computed in the experiments. The fourth column provides the average intersection over union (Jaccard similarity) between all MDS pairs. Next, the entropy, calculated based on the frequency of each actor's occurrence across all MDSs, is presented. The last column shows the profit achieved by applying the local improvement routine (Alg.~\ref{alg:driver_actor_selection}) to the dataset used in the study, compared to the greedy approach, in order to find an MDS. It is expressed as the difference between the sizes of the greedily obtained solution ($|D'|$) and the final MDS ($|\hat{D}|$), divided by the size of the former $(\frac{|D'|-|\hat{D}|}{|D'|})$, computed and averaged over the corresponding simulation runs.

\begin{table}[ht!]
    \centering
    \caption{Statistical properties of MDSs and the average size reduction (\textit{Avg. size red.}) of the initial $D'$ by the local improvement routine obtained for each network during the experiments. For column \textit{Avg. size} and \textit{Avg. size red.}, the standard deviation is reported in brackets.
    }
    \addtolength{\tabcolsep}{.3em}
    \begin{tabular}{l|rrrrr|r}
        \multicolumn{1}{c|}{Network} & \multicolumn{1}{c}{Size range} & \multicolumn{1}{c}{Avg. size} & \multicolumn{1}{c}{Unique} & \multicolumn{1}{c}{IoU} & \multicolumn{1}{c|}{Entropy} & \multicolumn{1}{c}{Avg. size red.} \\ \hline \hline
        er-2 & $[0.26, 0.27]$ & \scinot{0.27}{3}{-3} & $300 / 300$ & $0.42$ & $9.13$ & \scinot{0.15}{4}{-3} \\
        er-3 & $[0.34, 0.36]$ & \scinot{0.35}{4}{-3} & $300 / 300$ & $0.44$ & $9.41$ & \scinot{0.18}{5}{-3} \\
        er-5 & $[0.34, 0.36]$ & \scinot{0.35}{4}{-3} & $300 / 300$ & $0.39$ & $9.54$ & \scinot{0.20}{5}{-3} \\ \hline
        sf-2 & $[0.49, 0.50]$ & \scinot{0.50}{1}{-3} & $300 / 300$ & $0.88$ & $9.13$ & \scinot{0.06}{2}{-3} \\
        sf-3 & $[0.75, 0.76]$ & \scinot{0.76}{1}{-3} & $298 / 300$ & $0.97$ & $9.60$ & \scinot{0.05}{1}{-3} \\
        sf-5 & $[0.69, 0.70]$ & \scinot{0.70}{1}{-3} & $300 / 300$ & $0.92$ & $9.54$ & \scinot{0.09}{1}{-3} \\ \hline
        arxiv & $[1.00, 1.00]$ & \scinot{1.00}{0}{-0} & $1 / 300$ & \texttt{NAN} & \texttt{NAN} & \scinot{0.00}{0}{-0} \\
        aucs & $[0.28, 0.41]$ & \scinot{0.33}{2}{-2} & $77 / 300$ & $0.51$ & $5.01$ & \scinot{0.63}{7}{-3} \\
        ckmp & $[0.32, 0.35]$ & \scinot{0.33}{6}{-3} & $300 / 300$ & $0.62$ & $6.94$ & \scinot{0.31}{6}{-3} \\
        lazega & $[0.14, 0.17]$ & \scinot{0.15}{9}{-3} & $158 / 300$ & $0.31$ & $4.71$ & \scinot{0.25}{2}{-2} \\
        l2-course & $[0.20, 0.22]$ & \scinot{0.21}{1}{-2} & $149 / 300$ & $0.34$ & $4.36$ & \scinot{0.13}{2}{-2} \\
        timik & $[0.90, 0.91]$ & \scinot{0.91}{2}{-4} & $295 / 300$ & $0.99$ & $15.78$ & \scinot{0.00}{3}{-5} \\
    \end{tabular}
    \label{tab:mds_networks}
\end{table}

It is interesting to observe the differences between Erd\H{o}s-R\'{e}nyi and scale-free networks. The former can be controlled with significantly fewer driver actors than the latter. In both cases, nearly every simulation returned a different MDS. However, their similarity varies: Erd\H{o}s-R\'{e}nyi networks exhibit a much lower Jaccard similarity compared to scale-free networks. This suggests that driver actors in that group can be assigned in a more random manner than in scale-free graphs. 

With regards to real-world networks, MDS properties appear less regular. Therefore, in the following sections, we discuss the performance of the real graphs in relation to their structural similarity to both artificial types of networks. Nonetheless, the average $|\hat{D}|$ is comparable to that observed in scale-free (\textit{timik}) and Erd\H{o}s-R\'{e}nyi models (\textit{aucs}, \textit{ckmp}, \textit{lazega}, \textit{l2-course}). An interesting result was obtained for \textit{arxiv}. Due to its structural properties (on average, each actor is present in $1.91$ layers, whereas the total number of layers is $13$) and despite its relatively large size, the MDS for this network covers the entire set of actors. This case illustrates that, as in many aspects of network science, certain graph structures can present challenges that are difficult to address using general methods.

Entropy values appear to correlate with network size. For instance, the smallest network, \textit{l2-course}, exhibits the lowest entropy, while the largest, \textit{timik}, has the highest. This suggests that larger networks provide more diverse MDS configurations. Regarding the relationship between network layers and MDS size, we observe that an increase in the number of layers corresponds to a larger MDS. Finally, the number of unique MDSs is correlated with network size. For smaller networks (e.g., \textit{aucs}, \textit{lazega}, \textit{l2-course}), it is likely that all possible MDS configurations were sampled.

\subsubsection{Example of MDS alignment across artificial networks}

\begin{figure}[ht!]
    \centering
    \subfloat[MDS on an exemplary two-layer Erd\H{o}s-R\'{e}nyi network; $|A|=35, |\hat{D}|=4$]{
        \includegraphics[width=.93\linewidth]{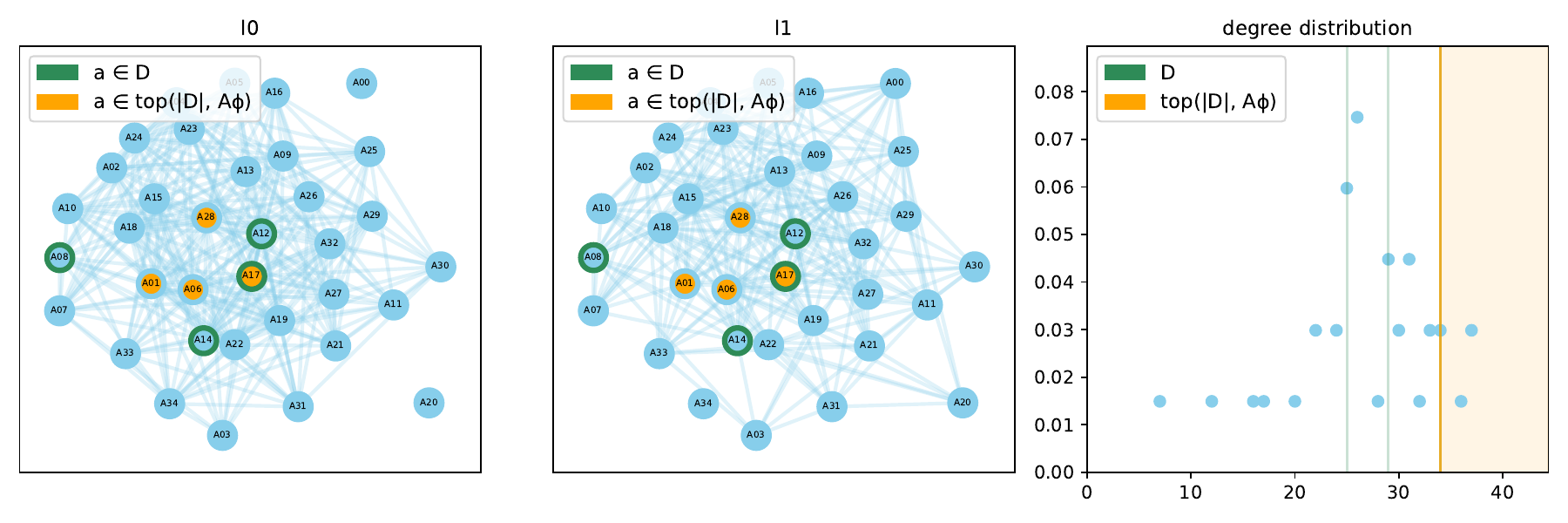}
        \label{subfig:er_mds}
    } \\
    \subfloat[MDS on an exemplary two-layer scale-free network; $|A|=35, |\hat{D}|=15$]{
        \includegraphics[width=.93\linewidth]{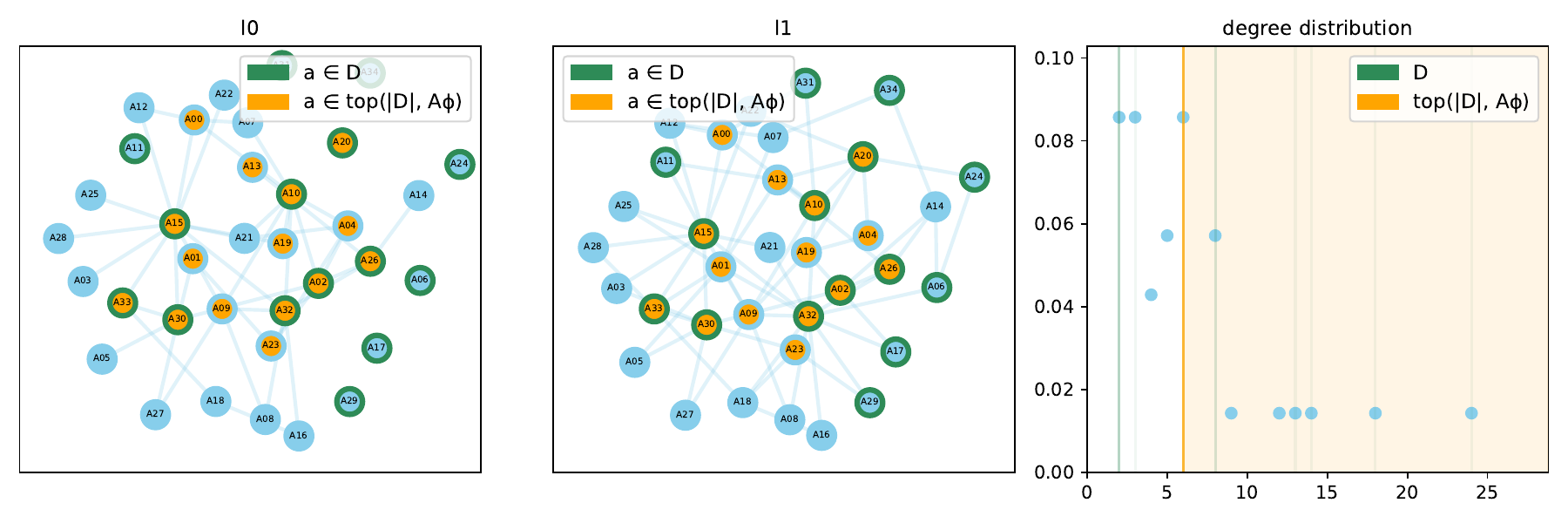}
        \label{subfig:pa_mds}
    }
    \caption{Typical MDS alignment with regard to the actors' degree. The first two plots depict layers of the given network, while the third shows the degree distribution. Green stripes indicate the degrees of actors belonging to the MDS; the orange area marks the degree range of a set composed of the highest-degree actors with the same size as the MDS. In Fig.~\ref{subfig:er_mds}, the MDS exhibits a much stronger correlation with the highest-degree actors compared to Fig.~\ref{subfig:pa_mds}, where the MDS is larger and includes both hubs and low-degree actors.
    }
    \label{fig:mds_vis}
\end{figure}

Differences in MDS's properties regarding Erd\H{o}s-R\'{e}nyi and scale-free models can be illustrated with Fig.~\ref{fig:mds_vis}. Since the degree distribution of the latter follows a power law, controlling the network requires selecting both the hubs (e.g., actors $A15$, $A32$) and outliers, which may have connections on only some layers (e.g., $A06$ or $A17$). Consequently, a larger number of actors must be selected to exert control over the network. On the other hand, Erd\H{o}s-R\'{e}nyi networks, which follow a Poisson distribution, exhibit a more even degree distribution. As a result, the MDS can be smaller (e.g., $|\hat{D}|=4$ compared to $|\hat{D}|=15$) and selected mostly from actors with the highest degree.

\subsubsection{Approximation of the globally optimal solution}

Finally, to assess the robustness of Alg.~\ref{alg:driver_actor_selection}, we analyze the extent to which it reduces the greedily obtained initial solution (as reported in the last column of Tab.~\ref{tab:mds_networks}). The greatest enhancement resulting from the local improvement routine was observed for the \textit{aucs} network, where the algorithm produced an MDS that was, on average, only $37\%$ the size of the initial solution. The only two networks for which no significant reduction in $D'$ was achieved are \textit{arxiv}, where MDS covers the entire network, and \textit{timik}, due to the timeout mechanism imposed to complete the computations within a reasonable time frame. Furthermore, the scale-free nature of this network suggests that the expected size of the MDS is inherently large. Nonetheless, one can conclude that Alg.~\ref{alg:driver_actor_selection} effectively reduces the size of the dominating set compared to the baseline approach.

We also sought to assess how closely the MDSs returned by Alg.~\ref{alg:driver_actor_selection} approximate globally optimal solutions; that is, whether the obtained minimal dominating sets are also minimum dominating sets. To this end, we executed a brute-force algorithm to exhaustively explore the solution space up to the minimal size of the MDSs found for a given network. The number of combinations to evaluate using such an approach is $\sum_{n=0}^{|\hat{D}|}\binom{|A|}{n}$.

Due to the computational complexity of this task, the evaluation was conducted on the three smallest networks --- \textit{l2-course}, \textit{aucs}, and \textit{lazega}. Since the number of sets to evaluate grows rapidly, we were not able to explore all possible solutions for bigger networks in a reasonable time. Nonetheless, the complete results for \textit{l2-course} indicate that Alg.~\ref{alg:driver_actor_selection} was able to return the minimum dominating set. For \textit{lazega}, we explored all combinations up to sets consisting of 8 actors; thus, recalling that $_{min}|\hat{D}|=10$ was obtained with the Local Improvement routine, we conclude that $|D| \in \{9, 10\}$, making Alg.~\ref{alg:driver_actor_selection} a relatively efficient approach. Finally, since the number of layers in \textit{aucs} was the largest among the three networks discussed, the best result obtained with the Local Improvement routine was $_{min}|\hat{D}|=17$. We were able to explore all combinations up to 9-element sets. Therefore, $|D| \in \{10, \dots, 17\}$.

\subsection{Role of MDS in seed selection}

To evaluate MDS's role in seed selection, we compared all obtained seed sets (i.e., from experiments where $|\hat{D}| \geq s$). They were grouped into pairs based on the protocol, budget size, experiment repetition, and network for both the baseline and MDS-filtered methods (activation thresholds were omitted, as they do not affect seeding). The Jaccard similarity was computed for each of these pairs. Finally, the similarity scores were averaged over each network type, seed selection method, and budget size.

\begin{table}[ht]
    \centering
    \caption{Similarities between seed sets obtained using the baseline methods and their MDS-backed variants, grouped by network type and budget size.
    }
    \addtolength{\tabcolsep}{.75em}

    \begin{tabular}{l|rrrrr}
    \multicolumn{6}{c}{Avg. $IOU(\phi, \phi_\kappa)$ --- Erd\H{o}s-R\'{e}ny networks} \\ \hline \hline
    \multicolumn{1}{c}{$s$} & \multicolumn{1}{c}{\textit{deg-c}} & \multicolumn{1}{c}{\textit{deg-cd}} & \multicolumn{1}{c}{\textit{nghb-1s}} & \multicolumn{1}{c}{\textit{nghb-sd}} & \multicolumn{1}{c}{\textit{random}} \\ \hline
     5 & $0.48$ $(0.08)$ & $0.50$ $(0.08)$ & $0.41$ $(0.07)$ & $0.42$ $(0.07)$ & $0.02$ $(0.02)$ \\
    10 & $0.41$ $(0.05)$ & $0.43$ $(0.04)$ & $0.37$ $(0.05)$ & $0.35$ $(0.04)$ & $0.05$ $(0.01)$ \\
    15 & $0.36$ $(0.03)$ & $0.39$ $(0.03)$ & $0.34$ $(0.03)$ & $0.33$ $(0.02)$ & $0.08$ $(0.02)$ \\
    20 & $0.34$ $(0.02)$ & $0.36$ $(0.02)$ & $0.32$ $(0.02)$ & $0.32$ $(0.02)$ & $0.11$ $(0.02)$ \\
    25 & $0.33$ $(0.02)$ & $0.34$ $(0.02)$ & $0.31$ $(0.02)$ & $0.31$ $(0.02)$ & $0.14$ $(0.02)$ \\
    30 & $0.31$ $(0.01)$ & $0.32$ $(0.02)$ & $0.30$ $(0.01)$ & $0.29$ $(0.01)$ & $0.17$ $(0.01)$ \\
    35 & $0.30$ $(0.01)$ & $0.31$ $(0.02)$ & $0.29$ $(0.01)$ & $0.29$ $(0.02)$ & $0.21$ $(0.01)$ \\
    \end{tabular} \\

    \begin{tabular}{l|rrrrr}
    \multicolumn{6}{c}{Avg. $IOU(\phi, \phi_\kappa)$ --- scale-free networks} \\ \hline \hline
    \multicolumn{1}{c}{$s$} & \multicolumn{1}{c}{\textit{deg-c}} & \multicolumn{1}{c}{\textit{deg-cd}} & \multicolumn{1}{c}{\textit{nghb-1s}} & \multicolumn{1}{c}{\textit{nghb-sd}} & \multicolumn{1}{c}{\textit{random}} \\ \hline
     5 & $0.49$ $(0.07)$ & $0.53$ $(0.08)$ & $0.47$ $(0.06)$ & $0.46$ $(0.07)$ & $0.03$ $(0.02)$ \\
    10 & $0.37$ $(0.03)$ & $0.38$ $(0.04)$ & $0.34$ $(0.03)$ & $0.35$ $(0.03)$ & $0.05$ $(0.02)$ \\
    15 & $0.31$ $(0.04)$ & $0.34$ $(0.02)$ & $0.30$ $(0.02)$ & $0.32$ $(0.02)$ & $0.08$ $(0.02)$ \\
    20 & $0.29$ $(0.03)$ & $0.32$ $(0.02)$ & $0.28$ $(0.03)$ & $0.30$ $(0.04)$ & $0.11$ $(0.02)$ \\
    25 & $0.27$ $(0.03)$ & $0.30$ $(0.03)$ & $0.27$ $(0.05)$ & $0.29$ $(0.06)$ & $0.14$ $(0.01)$ \\
    30 & $0.27$ $(0.04)$ & $0.29$ $(0.05)$ & $0.26$ $(0.06)$ & $0.30$ $(0.07)$ & $0.18$ $(0.01)$ \\
    35 & $0.26$ $(0.06)$ & $0.28$ $(0.06)$ & $0.27$ $(0.08)$ & $0.30$ $(0.08)$ & $0.21$ $(0.02)$ \\
    \end{tabular} \\

    \begin{tabular}{l|rrrrr}
    \multicolumn{6}{c}{Avg. $IOU(\phi, \phi_\kappa)$ --- real networks} \\ \hline \hline
    \multicolumn{1}{c}{$s$} & \multicolumn{1}{c}{\textit{deg-c}} & \multicolumn{1}{c}{\textit{deg-cd}} & \multicolumn{1}{c}{\textit{nghb-1s}} & \multicolumn{1}{c}{\textit{nghb-sd}} & \multicolumn{1}{c}{\textit{random}} \\ \hline
     5 & $0.57$ $(0.27)$ & $0.61$ $(0.23)$ & $0.56$ $(0.29)$ & $0.54$ $(0.29)$ & $0.04$ $(0.06)$ \\
    10 & $0.53$ $(0.25)$ & $0.55$ $(0.24)$ & $0.49$ $(0.28)$ & $0.51$ $(0.28)$ & $0.06$ $(0.05)$ \\
    15 & $0.52$ $(0.26)$ & $0.53$ $(0.25)$ & $0.49$ $(0.28)$ & $0.50$ $(0.28)$ & $0.08$ $(0.05)$ \\
    20 & $0.53$ $(0.27)$ & $0.55$ $(0.26)$ & $0.52$ $(0.28)$ & $0.54$ $(0.28)$ & $0.11$ $(0.04)$ \\
    25 & $0.57$ $(0.28)$ & $0.57$ $(0.27)$ & $0.56$ $(0.28)$ & $0.59$ $(0.27)$ & $0.14$ $(0.03)$ \\
    30 & $0.56$ $(0.29)$ & $0.57$ $(0.28)$ & $0.55$ $(0.29)$ & $0.59$ $(0.27)$ & $0.18$ $(0.04)$ \\
    35 & $0.74$ $(0.26)$ & $0.80$ $(0.20)$ & $0.75$ $(0.25)$ & $0.82$ $(0.18)$ & $0.21$ $(0.01)$ \\
    \end{tabular}

    \label{tab:mds_role}
\end{table}

Tab.~\ref{tab:mds_role} presents the outcomes. A higher Jaccard similarity suggests a less prominent role of MDS, whereas lower values indicate a greater impact of MDS on the given heuristic. The results indicate that the average similarity between seed sets across all cases is approximately $0.36$, demonstrating that MDS filtering consistently influences the seed selection process. However, we observe higher $IOU$ values for real networks, which is due to exceptionally large MDSs in the \textit{arxiv} and \textit{timik} networks. Apart from these cases, similarity values are accompanied by relatively small standard deviations, which supports the observed trend. As expected, the lowest similarity between seed sets is observed for the \textit{random} selection method, as it lacks a structured approach to choosing influential nodes. The remaining seed selection methods exhibit comparable IoU across different budget sizes, suggesting that MDS filtering affects them uniformly, thereby underscoring its universal applicability.

For both Erd\H{o}s-R\'{e}nyi and scale-free networks, the similarity between seed sets obtained with and without MDS filtering consistently decreases as the budget size increases (with the exception of \textit{random}). This trend suggests that the impact of MDS filtering increases as the seeding budget grows. For smaller budget sizes, where the influence of MDS is less significant (i.e., IoU values are higher), the similarity between baseline and MDS-filtered seed sets is greater for scale-free networks than for Erd\H{o}s-R\'{e}nyi graphs. This observation aligns with the findings presented in Tab.~\ref{tab:mds_networks}, which indicate that MDS exhibits higher variability in the latter network model. Consequently, there exist more possible MDS configurations that do not always align with the baseline ranking.

In real networks, trends reflect the diverse structural patterns present in these graphs, making them less straightforward to interpret. The markedly higher standard deviation values (by an order of magnitude) further confirm the greater variability in seed set similarities within this cohort. Nonetheless, even with such unfavourable networks as \textit{arxiv} and \textit{timik} included, the role of MDS filtering remains both visible and significant.

\subsection{Effectiveness of MDS in influence maximization}
\label{subsec:heatmaps}

\begin{figure}[hp!]
    \centering
    \captionsetup[subfloat]{labelformat=empty}
    \subfloat[]{
        \includegraphics[page=9, width=.235\linewidth]{heatmaps_aggregated.pdf}
        \label{subfig:er_and_heatmap_gain}
    }
    \subfloat[]{
        \includegraphics[page=10, width=.235\linewidth]{heatmaps_aggregated.pdf}
        \label{subfig:er_and_heatmap_auc}
    }
    \subfloat[]{
        \includegraphics[page=11, width=.235\linewidth]{heatmaps_aggregated.pdf}
        \label{subfig:er_or_heatmap_gain}
    }
    \subfloat[]{
        \includegraphics[page=12, width=.235\linewidth]{heatmaps_aggregated.pdf}
        \label{subfig:er_or_heatmap_auc}
    } \\ \vspace{-.9cm}
    \subfloat[]{
        \includegraphics[page=5, width=.235\linewidth]{heatmaps_aggregated.pdf}
        \label{subfig:sf_and_heatmap_gain}
    }
    \subfloat[]{
        \includegraphics[page=6, width=.235\linewidth]{heatmaps_aggregated.pdf}
        \label{subfig:sf_and_heatmap_auc}
    }
    \subfloat[]{
        \includegraphics[page=7, width=.235\linewidth]{heatmaps_aggregated.pdf}
        \label{subfig:sf_or_heatmap_gain}
    }
    \subfloat[]{
        \includegraphics[page=8, width=.235\linewidth]{heatmaps_aggregated.pdf}
        \label{subfig:sf_or_heatmap_auc}
    } \\ \vspace{-.9cm}
    \subfloat[]{
        \includegraphics[page=1, width=.235\linewidth]{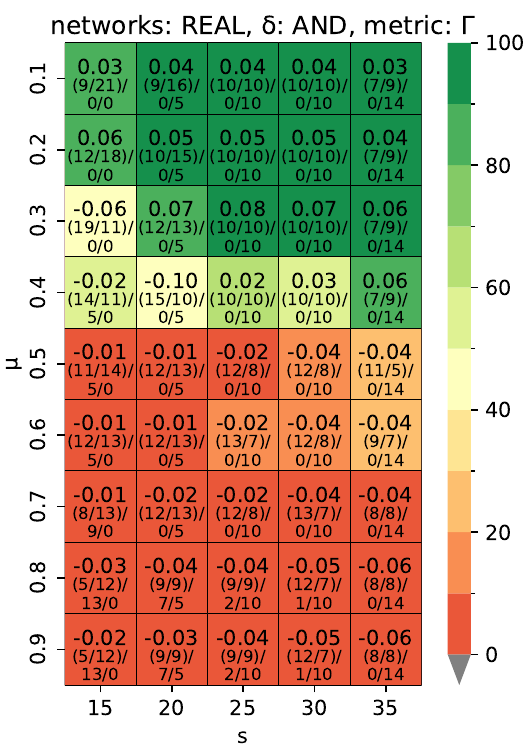}
        \label{subfig:real_and_heatmap_gain}
    }
    \subfloat[]{
        \includegraphics[page=2, width=.235\linewidth]{heatmaps_aggregated.pdf}
        \label{subfig:real_and_heatmap_auc}
    }
    \subfloat[]{
        \includegraphics[page=3, width=.235\linewidth]{heatmaps_aggregated.pdf}
        \label{subfig:real_or_heatmap_gain}
    }
    \subfloat[]{
        \includegraphics[page=4, width=.235\linewidth]{heatmaps_aggregated.pdf}
        \label{subfig:real_or_heatmap_auc}
    }
    \vspace{-.9cm}
    \caption{Improvement of MDS filtering across network types and spreading regimes, measured with $\Gamma$ and $\Lambda$. Each tile shows six values: the \textbf{upper value} is the mean difference between the baseline and MDS-filtered variant ($[-1, 1]$) across feasible experiments; the \textbf{middle row} presents the count of feasible experiments, split into those with a difference greater (\textbf{left}) or less than/equal to $0.01$ (\textbf{right}); the \textbf{lower row} shows unfeasible experiments where diffusion couldn't start (due to an overly strict spreading regime, \textbf{left}) or where $|\hat{D}| < s$ (\textbf{right}). The \textbf{tile colour} represents the percentage of feasible simulations where the MDS-filtered variant outperformed the baseline, computed over feasible simulations with an absolute difference greater than $0.01$.
    }
    \label{fig:heatmaps}
\end{figure}

Since this study aims to evaluate the general usability of MDS for seed set selection, we have deliberately omitted the influence of specific seed selection methods. To keep the report concise, we have also aggregated the results across different network types. Figure~\ref{fig:heatmaps}, which is discussed below, presents a comprehensive summary of our experimental results in a structured format. Readers interested in a more detailed analysis are encouraged to explore the repository.

\subsubsection{Feasibility of the experiments}

We begin our analysis with a quantitative summary of the experiments, focusing on the second and third rows of each tile in the heatmap. The former represents the number of experiments that were executed correctly, while the latter provides insights into the cases where simulations failed. Specifically, failures occurred either due to overly restrictive diffusion parameters (left value) or when the MDS was too small to select a seed set with it (right value). It is important to note that the corresponding values of failed experiments should be identical for pairs of $\Gamma$ and $\Lambda$ heatmaps that represent the same dataset fold and applied protocol. Moreover, the number of cases where MDS was too small to trigger diffusion is independent of $\mu$ and $\delta$. Thus, these values should be the same for all experiments conducted for the given $s$ and the network type.

\subparagraph{Failures due to overly demanding parameters.}

In Erd\H{o}s-R\'{e}nyi networks under the $AND$ protocol, failures occur for $\mu \geq 0.7$, with their number increasing as $\mu$ grows. Conversely, as the budget increases, the number of failures decreases. In the $OR$ regime, all experiments were feasible. A similar pattern is observed in scale-free networks for $\delta = AND$; however, the reduction in failed experiments due to an increasing budget occurs at a higher rate. For $\delta = OR$, all experiments were feasible. Finally, for real networks and spreading under the $AND$ protocol, the region of overly strict parameters corresponds to $\mu \geq 0.4$ or a budget of $s \leq 30$. In the $OR$ regime, failures occur for $\mu \geq 0.5$ or $s \leq 10$. 

As expected, the number of failed experiments depends on both $\mu$ and $s$. The highest failure rates occur for large $\mu$ and small $s$, while relaxing at least one of these parameters increases the likelihood of triggering a spread. Compared to artificial networks, the subset of real networks in our study poses greater challenges in initiating the diffusion process.

\subparagraph{Failures due to insufficient size of MDS.}

In Erd\H{o}s-R\'{e}nyi networks, for budgets $s \geq 30$, failures resulting from an insufficiently large MDS accounted for $33.33\%$ of the cases. In contrast, in scale-free networks, the MDS was consistently large enough to select a seed set that matched the budget, and no such failures were observed. For real networks, failures of this nature began to manifest for budgets $s \geq 20$, with the frequency of these failures increasing as the budget size grew. This trend aligns with the analysis of the MDS size, as presented in Tab.~\ref{tab:mds_networks}.

\subparagraph{Feasible experiments.}

The conditions that affect the number of feasible experiments (represented by the sum of the middle row in the tiles of the heatmaps) are the opposite of those described in the previous subsection. Specifically, a smaller threshold makes it easier to trigger the diffusion. A similar effect is observed with $s$, although here we encounter an obstacle in the form of the MDS size, which may be smaller than $s$ and thus prevent the simulation from being executed.

\newgeometry{textwidth=170mm}

\restoregeometry

\subsubsection{Average performance differences.}

This part of the analysis examines the average difference (the top value on each heatmap's tile and denoted as $\Delta$) between experiments with and without the MDS filtering operation, using both metrics employed.

\subparagraph{Erd\H{o}s-R\'{e}nyi networks.}

Results for spreading triggered by the $AND$ protocol indicate that, in general, the differences for $\Gamma$ tend to increase with the budget across experiments with a similar $\mu$. For small thresholds, the average differences in spreading effectiveness favour MDS-backed methods, while for larger $\mu$, baseline methods tend to prevail. Similar trends are observed for $\Delta \Lambda$. For $\delta=OR$, $\Delta \Gamma$ shows no clear trends. The average difference fluctuates around zero and is generally smaller than for $AND$. For the diffusion dynamics, the trends mirror those seen for $\Delta \Gamma$.

\subparagraph{Scale-free networks.}

For the $AND$ protocol, significant differences are observed for $\Gamma$, with values reaching up to $0.26$. As the budget increases, so does the difference, and for small thresholds, MDS-backed methods outperform the baseline. However, for larger thresholds, baseline methods prevail. For $\Delta \Lambda$, similar trends to $\Delta \Gamma$ are observed, but the values are slightly smaller. For the $OR$ protocol, $\Delta \Gamma$ shows no strong trends. Positive values (favouring MDS) are very small and close to $0$, while negative values exhibit a larger magnitude. For $\Delta \Lambda$, most values are negative, suggesting that using MDS even slows down the diffusion.

\subparagraph{Real networks.}

For the $AND$ protocol, $\Delta \Gamma$ values are smaller than those for scale-free networks but usually larger than those in observed Erd\H{o}s-R\'{e}nyi graphs. For small $\mu$, MDS-backed methods tend to outperform the baseline, though less prominently than in scale-free networks, while for larger thresholds, baseline methods prevail. However, the trend of increasing differences with the budget is not so visible as in artificial graphs. A similar pattern holds for $\Delta \Lambda$. For $\delta = OR$, $\Delta \Gamma$ values are scattered around $0$ with no clear trends, similar to Erd\H{o}s-R\'{e}nyi networks. For $\Delta \Lambda$, values are more often negative than for $\Delta \Gamma$, but no strong trends emerge as well.

\subsubsection{Seed set improvement by MDS}

The next part of the analysis in Fig.~\ref{fig:heatmaps} concerns the percentage of feasible simulations where MDS-backed methods "significantly" outperform their baseline versions, as indicated by colour. This comparison is intended to serve as a guideline for identifying conditions under which incorporating MDS into the seed selection process enhances diffusion, has no substantial impact, or even hinders the spreading process. In the following section, we provide a detailed discussion of each heatmap. Before that, we explain the methodology used to construct them.

For the purpose of analysis, we define a significant difference as an absolute difference according to $\Gamma$ or $\Lambda$ greater than $0.01$ ($1$ percentage point) between simulations conducted on the same network and under the same spreading regime but with seed sets selected using either the baseline method ($\phi$) or its MDS-backed variant ($\phi_\kappa$). The threshold of $0.01$ was chosen arbitrarily by the authors to exclude cases where the difference is negligible, i.e., when MDS neither significantly boosts nor disrupts diffusion.

To quantify the number of significantly different cases among all experiments, we split the count of feasible simulations into two values, represented in the second row of each tile: the left-hand value denotes the number of significantly different cases, while the right-hand value corresponds to cases where the difference is insignificant. The percentage of experiments in which $\phi_\kappa$ outperforms $\phi$ (i.e., represented by the tile colour) is computed using the former value as the denominator and, as the numerator, the count of significantly different cases where diffusion triggered with $\phi_\kappa$ was more effective than with $\phi$, according to the given metric.

Finally, it is important to note that the scale is symmetrical. The greener the tile, the more frequently MDS improves the seed selection process, whereas the less green it is, the more frequently the baseline method prevails. If no significantly different cases are observed for a given combination of network type, $s$, and $\mu$, the corresponding tile is displayed in grey. Naturally, MDS is not a favourable choice in such scenarios, as it provides no benefit while increasing the computational complexity.

\subparagraph{Erd\H{o}s-R\'{e}nyi networks.}

For the $AND$ protocol, MDS improves the overall effectiveness of diffusion in cases with small thresholds. The larger the budget, the greater the number of cases with a significant difference, and the average $\Delta \Gamma$ increases accordingly. We observe a transition boundary running diagonally across $\mu=0.4, s=25$, below which, regardless of $s$, the baseline methods remain more effective or the gain brought by MDS is insignificant. With regard to the spreading dynamics ($\Lambda$), we also observe a positive impact of MDS for diffusion with a small $\mu$ or a large $s$. Here, the transition between the region where baseline methods prevail is smoother, but in all cases with $\mu \geq 0.4$, classic approaches dominate.

For the $OR$ protocol, a different trend emerges for both metrics. MDS demonstrates its superiority in two regions corresponding to the extreme values of $\mu$. Between them, we identify a zone where baseline methods dominate, stretching diagonally from low values of both $s$ and $\mu$ to high values of these parameters. For $\Gamma$, this trend is compressed by the region of insignificantly differing runs, and MDS exhibits stronger dominance for $\mu \geq 0.8$ and $10 \leq s \leq 20$. For $\Lambda$, the aforementioned trend is more pronounced, with the region of the most significant improvement brought by MDS resembling that observed for $\Gamma$.

\subparagraph{Scale-free networks.}

Under the $AND$ protocol, scale-free networks exhibit the greatest improvement in spreading across both metrics. As in Erd\H{o}s-R\'{e}nyi graphs, experiments with $\mu > 0.4$ favour baseline methods, whereas below this threshold, MDS significantly enhances diffusion, both in terms of the proportion of superior cases and the average $\Delta \Gamma$. Notably, in this region, most feasible simulations show a substantial difference in $\Gamma$ between runs triggered by seed sets selected with baseline and MDS-backed methods, further reinforcing the advantage of the latter. A similar trend is observed for $\Lambda$, though the transition between MDS- and baseline-dominated regions appears rougher and more diagonally oriented, with $\Delta\Lambda$ generally lower than $\Delta\Gamma$.

For the $OR$ protocol, the heatmaps resemble those of Erd\H{o}s-R\'{e}nyi networks, forming a diagonal zone where baseline methods dominate. This region is broader for $\Lambda$ and narrower for $\Gamma$. Although the $\Gamma$ heatmap contains more green-coloured tiles, the improvement brought by MDS remains limited. In many cases, the observed advantage is deceptive --- the number of significantly differing results is small (at most $4$ out of $15$), and $\Delta\Gamma$ does not exceed $0.02$. Moreover, MDS-backed methods occasionally disrupt diffusion, as seen for $\mu=0.7, s=10$ or $\mu=0.8, s=25$. The results for spreading dynamics also support this finding, providing evidence that MDS does not improve diffusion in scale-free networks under the $OR$ protocol.

\subparagraph{Real networks.}

Heatmaps for the $AND$ protocol closely resemble those obtained for scale-free networks. However, the values of $\Delta \Gamma$ and $\Delta \Lambda$ are generally more aligned with those observed in Erd\H{o}s-R\'{e}nyi graphs, meaning the results are more concentrated around zero. The diagonal transition boundary between regions dominated by MDS and baseline methods appears more vertical and less abrupt.

In the $OR$ protocol, the correlation between both heatmaps is weaker than in the $AND$ case. Nevertheless, the same pattern seen in Erd\H{o}s-R\'{e}nyi networks emerges, where zones of strong MDS dominance are divided by a diagonal region stretching from small $\mu$ and $s$ to large values of both parameters. For $\Gamma$, seed selection methods incorporating MDS filtering are favoured in most cases, but the average improvement remains modest, not exceeding $0.04$. In contrast, results for $\Lambda$ are more evenly distributed, with the proportion of significantly feasible cases where MDS-backed methods outperform their baseline counterparts fluctuating around $50\%$.

\subsection{Applicability of MDS in relation to the properties of scale-free networks}
\label{subsec:sf_followup}

As shown in Fig.\ref{fig:heatmaps}, the usefulness of incorporating the MDS into the seed selection process depends on both the network type and the spreading conditions. The results indicate that scale-free networks benefit most from control-based approaches. Given the ubiquity of such structures in real-world systems~\cite{barabasi2009sfnetworks}, this finding is particularly relevant. However, like any synthetic model, the scale-free model has parameters that introduce additional factors. In the aforementioned experiments, we intentionally omitted these to avoid adding further variables to the evaluation. Nonetheless, since the results for scale-free networks demonstrated strong potential for MDS-backed seed selection, we conducted a follow-up experiment to assess whether this positive effect persists in these networks regardless of their configuration parameters.

\subsubsection{Experiment scope}

The evaluated properties of the network model had to be closely aligned with the \texttt{multinet} \cite{magnani2021analysis} library, which was employed in the previous parts of the study. This framework is designed to generate artificial multilayer networks and organizes configuration parameters into two groups: those governing intralayer properties, which determine the growth of graphs representing specific relations, and those defining interlayer dependencies, reflecting relations among the modelled layers. Two parameters from the configuration model were selected for evaluation, forming the basis for two cohorts of simulations. The first concerned the number of actors in the network. The second related to the number of initial nodes per layer from which the network growth process begins --- these have a substantial influence on the emergence of hubs within each layer. All remaining parameters were held constant and set to typical default values.

The initial configuration setup of the network generator assumed $1,000$ actors represented across three layers, each constructed using the preferential attachment model~\cite{sf_model} with identical parameters: $m_0=m=6$. Regarding interlayer dependencies, the probability of adding an internal edge was set to $0.7$, the probability of randomly importing an edge from another layer was $0.2$, and the probability that no action occurred in a given growth step (i.e., the network grew more slowly) was $0.1$. The dependency matrix, which served as the probability distribution for selecting the source layer from which to import an edge, was homogeneous, with values equal to $1/(|L|-1)$, except for the diagonal entries, which were set to zero. Finally, the number of growth steps was set as $|A|-m_0$. In addition, for each utilized configuration setup, $20$ instances of networks were created and used in simulations to mitigate the randomness effect.

As for the other conditions, we selected a subset of the full parameter space presented in Tab.~\ref{tab:parameters}, which appeared favourable in terms of the applicability of MDS to influence maximization in scale-free networks (see Fig.~\ref{fig:heatmaps}). All five seed selection heuristics were employed in two variants: the baseline and the MDS-backed versions. Only one spreading scenario was considered: a budget of $30\%$ of actors, an activation threshold of $0.2$, and the $AND$ protocol function. Each simulated case was repeated $30$ times to mitigate stochastic effects, such as those arising from obtaining the MDS.

\subsubsection{Number of actors}
\label{subsubsec:sf_followup_actors}

The size of a network plays an important role in shaping its properties. To assess whether it also affects the effectiveness of using seed set drawn from the MDS, we evaluated five series of scale-free networks with varying numbers of actors: $500$, $750$, $1000$, $1250$, and $1500$ (in total $1,000$ networks).

\begin{table}[ht!]
    \centering
    \caption{Influence boost by incorporating MDS into the seed selection process across scale-free networks with varying numbers of actors ($|A|$), according to the overall spread ($\Gamma$) or dynamics ($\Lambda$).
    }
    \addtolength{\tabcolsep}{.75em}
    \begin{tabular}{llllll}
    & \multicolumn{1}{c}{$|A|=500$} & \multicolumn{1}{c}{$|A|=750$} & \multicolumn{1}{c}{$|A|=1000$} & \multicolumn{1}{c}{$|A|=1250$} & \multicolumn{1}{c}{$|A|=1500$} \\ \hline \hline
    $\Delta\Gamma$ & $0.26(0.04)$ & $0.27(0.03)$ & $0.26(0.03)$ & $0.27(0.03)$ & $0.24(0.02)$ \\
    $\Delta\Lambda$ & $0.14(0.04)$ & $0.13(0.03)$ & $0.13(0.03)$ & $0.15(0.03)$ & $0.12(0.02)$ \\
    \end{tabular}
    \label{tab:sf_var_actors}
\end{table}

Tab.~\ref{tab:sf_var_actors} presents the experimental results, expressed as the mean difference in diffusion performance, measured in terms of overall spread ($\Gamma$) and dynamics ($\Lambda$), between corresponding runs using MDS-backed and baseline seed selection heuristics. As can be observed, both metrics are similar across the network cohorts, indicating that the applicability of MDS does not depend on the number of actors in the considered spreading scenario. Moreover, the improvement in spreading is consistent and in line with the results depicted in Fig.~\ref{fig:heatmaps}.

\subsubsection{Number of initial nodes}

The second parameter of the scale-free model we evaluated was $m_0$. It governs the number of initial nodes used in the first step of the graph generation process and, consequently, is closely related to the expected number of hubs. As we consider multilayer networks, $m_0$ could take different values in each layer. However, to facilitate more straightforward conclusions, we chose to keep it uniform across all layers. The number of links created by a newly added node ($m$) was also fixed, with $m=m_0$.

\begin{table}[ht!]
    \centering
    \caption{Influence boost by incorporating MDS into the seed selection process across scale-free networks with varying numbers of initial nodes ($m_0$), according to the overall spread ($\Gamma$) or dynamics ($\Lambda$).
    }
    \addtolength{\tabcolsep}{.83em}
    \begin{tabular}{llllll}
    & \multicolumn{1}{c}{$m_0=2$} & \multicolumn{1}{c}{$m_0=4$} & \multicolumn{1}{c}{$m_0=6$} & \multicolumn{1}{c}{$m_0=8$} & \multicolumn{1}{c}{$m_0=10$} \\ \hline \hline
    $\Delta\Gamma$ & $0.17(0.03)$ & $0.23(0.03)$ & $0.26(0.03)$ & $0.29(0.04)$ & $0.29(0.03)$ \\
    $\Delta\Lambda$ & $0.14(0.03)$ & $0.16(0.03)$ & $0.13(0.03)$ & $0.15(0.03)$ & $0.11(0.03)$ \\
    \end{tabular}
    \label{tab:sf_var_hubs}
\end{table}

The results are presented in Tab.\ref{tab:sf_var_hubs}, following the format used in Sec.\ref{subsubsec:sf_followup_actors}. One can note that they are broadly consistent with the trends observed in Fig.~\ref{fig:heatmaps}, although interpretation is less straightforward due to one anomaly. From the perspective of $\Gamma$, the benefit of incorporating the MDS, while consistently substantial, diminishes as $m_0$ decreases. This may be because, for low values of this parameter, hubs are more pronounced, and the baseline seed selection heuristics, which largely rely on centrality measures, can already identify seed nodes effectively. As the number of hubs increases, however, the role of the MDS becomes more prominent, as it helps focus on the most influential hubs.

\section{Conclusions}
\label{sec:conclusions}

In this work, we examined the impact of using MDS for seed selection in a problem of influence maximization in multilayer networks under the Linear Threshold Model. We conducted a comprehensive analysis across five rank-refining seed selection heuristics, eleven networks, and a wide range of diffusion regimes. Additionally, we adapted the local improvement algorithm for finding MDS to the multilayer network setting.

\subsection{Key Findings}

After analyzing the results, we conclude that using MDS in seed selection process benefits diffusion under low activation thresholds (up to $\mu=0.4$) and the $\delta=AND$, regardless of the network type. In real-world scenarios, this spreading regime may correspond to situations where convincing someone requires fewer acquaintances, but these connections must span all layers (see the election example in Sec.~\ref{sec:intro}). Higher seeding budgets further amplify this effect, as supported by Tab.~\ref{tab:mds_role}, which shows that the role of MDS becomes more significant with increasing $s$. The underlying reason is that MDS distributes seeds more widely across the network, including its periphery, whereas baseline methods concentrate them in the core. At lower $\mu$, this ensures faster coverage of the entire network. Baseline methods, confined to the core, lack seeds in the periphery. As the threshold increases, MDS-backed methods cease to be effective sooner, as their influence is split between the core and periphery.

Under the $OR$ protocol, where spreading is inherently easier, the effects observed in $AND$ are far less pronounced. A seed set strong on a single layer is often sufficient to drive diffusion across the entire network, which baseline methods achieve more effectively. However, when traditional approaches fail under high thresholds, MDS can offer an advantage. In situations where finding seed set through standard methods becomes challenging, MDS facilitates diffusion that would otherwise not occur.

We also observe that scale-free networks are the most susceptible to diffusion improvement with MDS. This stems from the presence of both hub and peripheral nodes in MDS (Fig.~\ref{fig:mds_vis}). Unlike Erd\H{o}s-R\'{e}nyi networks, which distribute connections more evenly, scale-free structures benefit from seed sets that span both the core and periphery, making MDS-backed methods more effective in triggering diffusion. Moreover, the role of MDS in seed selection is the most pronounced for this networks compared to the others (Tab.~\ref{tab:mds_role}) and persists regardless of its parameters (Tab.~\ref{tab:sf_var_actors}, Tab.~\ref{tab:sf_var_hubs}). In contrast, Erd\H{o}s-R\'{e}nyi networks are less suitable for MDS-based approaches due to their degree distribution, which results in MDS being more randomized and typically too small (see Tab.~\ref{tab:mds_networks}) to meet the budget requirements.

For the real networks analyzed, MDS sizes (Tab.~\ref{tab:mds_networks}) suggest this subset includes both networks resembling scale-free and Erd\H{o}s-R\'{e}nyi models, making it the most representative of real-world scenarios. As shown in Fig.~\ref{fig:heatmaps}, MDS offers the greatest benefit under the $AND$ protocol with lower activation thresholds, improving overall coverage and diffusion dynamics. However, the example of \textit{arxiv} shows that, in some cases, using the MDS is meaningless when it covers the entire set of actors.

Finally, despite effectiveness, using MDS comes with additional computational costs (Sec. \ref{subsubsec:mds_ssm}) that need to be taken into consideration when choosing a seed selection approach. Nonetheless, in some cases, the increase in $\Lambda$ or $\Gamma$ might be big enough to justify the additional time needed to compute MDS.

\subsection{Limitations of the Study and Next Steps}

This study includes a diverse set of networks, comprising both real-world examples and representatives of scale-free and Erd\H{o}s-R\'{e}nyi graphs. However, a more in-depth evaluation of the artificial models employed may provide additional insight, particularly in the case of the latter model, which was omitted owing to discouraging results. Future work may also explore the applicability of our findings to alternative graph structures, such as small-world networks. Similarly, the impact of adopting different spreading models, such as the Independent Cascade, could be investigated. Addressing these limitations thus presents a promising direction for further research.

Another possible avenue for future work is to explore alternative interpretations of the control problem in multilayer networks. Since this graph model is built upon the idea of capturing multiple types of relations among a set of agents, Def.~\ref{def:ds} was formulated such that an actor must be dominated in all (as opposed to just any) layers of the network in order to be considered controlled. Nevertheless, one can conceive an alternative formulation in which an actor is regarded as dominated if it is linked to a dominating actor in at least one layer. The MDS derived under this relaxed condition would typically be smaller than those obtained using the stricter formulation adopted in this work, and the resulting influence dynamics may also differ.

While the steps outlined here are only examples of potential directions for deeper exploration of the problem undertaken in this work, the primary conclusion remains clear: if one seeks to persuade a wide audience, particularly those whose conviction is easily swayed but must be thoroughly convinced across all aspects of their lives, starting from MDS is a good choice.

\section*{Acknowledgment}

This work was supported by the Australian Research Council [grant no. DP190101087]; the National Science Centre, Poland [grant no. 2022/45/B/ST6/04145]; the Polish Ministry of Science and Higher Education programme “International Projects Co-Funded”; and the EU under the Horizon Europe [grant no. 101086321]. Views and opinions expressed are those of the authors and do not necessarily reflect those of the funding agencies.

\printbibliography

\end{document}